\documentstyle[12pt,a4,a4wide,epsfig]{article}
 
\def\be{\begin{equation}}
\def\ee{\end{equation}}

\def\ra{\rightarrow}
\def\be{\begin{equation}}
\def\ee{\end{equation}}
\def\gs{\mathrel{
   \rlap{\raise 0.511ex \hbox{$>$}}{\lower 0.511ex \hbox{$\sim$}}}}
\def\ls{\mathrel{
   \rlap{\raise 0.511ex \hbox{$<$}}{\lower 0.511ex \hbox{$\sim$}}}}
\newcommand{\obb}{0\mbox{$\nu\beta\beta~$}}
\newcommand{\onbb}{neutrinoless double beta decay}

\newcommand{\dma}{\mbox{$\Delta m^2_{\rm A}~$}}
\newcommand{\dms}{\mbox{$\Delta m^2_\odot~$}}
\newcommand{\ba}{\begin{array}{c}}  
\newcommand{\bad}{\begin{array}{ccc}}
\newcommand{\bea}{\begin{equation} \begin{array}{c}}
\newcommand{\eea}{ \end{array} \end{equation}}
\newcommand{\ea}{\end{array}}
\newcommand{\D}{\displaystyle} 
\newcommand{\meff}{\mbox{$\langle m \rangle~$}}

\newcommand{\lsim}{{\;\raise0.3ex\hbox{$<$\kern-0.75em\raise-1.1ex\hbox{$\sim$}} 
\;}} 
\newcommand{\gsim}{{\;\raise0.3ex\hbox{$>$\kern-0.75em\raise-1.1ex\hbox{$\sim$}} 
\;}} 
 
\begin{document} 
\title{
\hfill { \bf {\small DO--TH 01/20}}\\
\hfill { \bf {\small hep-ph/0201052}}\\ \vskip 1cm  
\bf Leptogenesis and low energy observables in left--right 
symmetric models}
{\it 
\author{ K.R.S.~Balaji\thanks{Email address: 
balaji@zylon.physik.uni-dortmund.de}, $\;\;$
Werner Rodejohann\thanks{Email address: 
rodejoha@xena.physik.uni-dortmund.de}\\ \\
{\normalsize \it Lehrstuhl f\"ur Theoretische Physik III,}\\ 
{\normalsize \it Universit\"at Dortmund, Otto--Hahn--Str. 4,}\\ 
{\normalsize \it 44221 Dortmund, Germany}}} 
\date{}
\maketitle
\thispagestyle{empty}

\begin{abstract} 
  In the context of left--right symmetric models 
we study the connection of leptogenesis and low energy parameters such as 
neutrinoless double beta decay and leptonic $CP$ violation. 
Upon imposition of a unitarity constraint, 
the neutrino parameters are significantly restricted and the Majorana 
phases are determined within a narrow range, depending on the kind 
of solar solution. One of the Majorana phases gets determined to a good
accuracy and thereby the second phase can be probed from the 
results of neutrinoless double beta decay experiments. 
We examine the contributions of the solar and atmospheric 
mass squared differences to the asymmetry and find that in general the 
solar scale dominates. In order to let the atmospheric scale dominate, 
some finetuning between one of the Majorana phases and the Dirac $CP$ phase 
is required. In this case, one of the Majorana phases is 
determined by the amount of $CP$ violation in oscillation experiments.
\end{abstract} 
 
\vspace{0.2cm} 


\newpage 
 
\section{Introduction} 
 
In the last few years, there has been mounting evidence for physics
beyond the standard model coming from the leptonic sector of the model. 
In particular, the muon up--down asymmetry, as declared by the 
SuperKamiokande collaboration, has given compelling evidence for neutrino
mass and mixing \cite{sk}. 
Measurements of the solar neutrino fluxes by several 
experiments \cite{solarexp} have also provided convincing experimental 
signatures for oscillations. Recently, the first results from the SNO 
experiment \cite{SNO} have substantiated 
the existence of neutrino oscillations 
among active flavors involving $\nu_e$ from the Sun. In parallel, an 
interesting problem in cosmology --- which could have its solution from the
particle physics sector --- is the issue of resolving a tiny 
baryon asymmetry in the universe \cite{basymm}. 
To recapitulate, the explanation of this 
asymmetry requires satisfying the 
three Sakharov conditions, one of them being the 
presence of $CP$ violation \cite{sakharov}.
As it is well known, within the standard model, $CP$ violation is explained
through a phase in the CKM matrix 
and turns out to produce a baryon asymmetry far too below \cite{smcp} the 
observed value \cite{citeYB} and thus additional inputs are required. 
For instance, the leptogenesis mechanism \cite{leptogenesis} 
can produce a baryon asymmetry 
through the out--of--equilibrium decay of heavy right--handed Majorana 
neutrinos in the early universe. Courtesy of the see--saw 
mechanism \cite{seesaw}, these right--handed Majorana neutrinos 
also produce small masses for the light left--handed neutrinos as 
indicated by neutrino experiments. In addition, most viable neutrino models 
with large mixings \cite{smbarr} produce 
Majorana mass terms which break the $B-L$ quantum number by two 
units and it is to be noted that if the conservation of this quantum 
number is assumed, the explanation of the baryon asymmetry is hard.\\

Therefore, the presence 
of heavy right--handed Majorana masses can be useful to explain both, the 
smallness of neutrino masses in oscillation experiments and the baryon 
asymmetry of the universe. This connection has been analyzed in several 
recent papers \cite{others1,JPR1,JPR2,ichAPP}.
It would be fair to say, that, from all of these observations, 
there seems to be a definite indication of new physics interplay 
between cosmology and particle physics, and leptogenesis could be one
viability.\\

 A relevant issue to this 
subject is to examine possible low energy signatures of 
leptogenesis. Among them are $CP$ violation in 
oscillation experiments and the value of the effective electron neutrino 
mass, \meff$\!\!$, as measured in \onbb{} $(0\nu\beta\beta)$. In fact, 
given that in most 
models the heavy Majorana neutrinos are too heavy to be produced at 
realistic collider energies, observation of low energy $CP$ violation 
and lepton number violation might be the only possibility to 
validate leptogenesis. It might even be possible to distinguish 
different models through these additional observables \cite{JPR2,ichAPP}. 
In this paper, we consider 
leptogenesis in left--right 
symmetric (LR) models\footnote{See e.g.\ \cite{zhang} 
for the possibility of 
baryogenesis in left--right models.} \cite{JPR1,JPR2,ichAPP}. 
We are motivated by the 
simplicity of the model which offers us to relate the left-- and
right--handed sectors of the theory due to the symmetry. This choice reduces 
the ambiguities which arise due to the unknown right--handed sector of the 
theory. In this model, one finds that for a specific choice of the 
Dirac mass matrix, the baryon asymmetry is 
proportional to the lightest mass eigenvalue. When we impose a 
unitarity constraint on this mass, we observe that rather stringent 
constraints on the low energy parameters follow, especially regarding the 
yet unknown phases in the leptonic mixing matrix. Subsequently, in a limiting
two flavor case, we can relate one of the 
Majorana phases to the solar mass squared difference and also set useful
lower bounds on the neutrino parameters.\\

Our paper is organized as follows. In the next section, we give the basic 
formalism of the see--saw mechanism in LR models. 
Section III deals with the leptogenesis mechanism and the results for 
the baryon asymmetry in the LR model. 
In Section IV, we apply a unitarity constraint to the 
light Majorana mass, which is used to constrain the low energy parameters. 
A lot of our analysis can be 
complemented with future solar neutrino 
experiments which will try to pin down the specific solar solution. 
We conclude in Section V by summarizing the main results.\\

\section{\label{sec:form}Basic formalism in left--right symmetric theories}
We begin by reviewing the known results for neutrino mass in LR models
\cite{JPR1,JPR2}. The see--saw mechanism follows in models where the fermionic
sector of the standard model is extended with massive  
right--handed singlet (under $SU(2)_L$ group action) neutrinos with mass 
of the order 
of $10^{10}$ GeV or heavier. The decoupling of such heavy mass states from the
active left--handed sector can result in a small Majorana mass. 
In LR symmetric theories this decoupling results in a mass term of the form
\be \label{mnu}
m_\nu = M_L - \tilde{M}_D \, M_R^{-1} \, \tilde{M}_D^T~.
\ee 
In (\ref{mnu}), $\tilde M_D$ and $M_R$ denote the Dirac and the heavy 
right--handed Majorana neutrino mass matrices, respectively. 
This is to be contrasted with the 
conventional form where the left--handed mass matrix $M_L$ is absent 
and hence does 
not contribute to the light neutrino mass. 
The presence of $M_L$ is required in order to maintain the LR 
symmetry. The matrix in (\ref{mnu}) can
be diagonalized in the usual way with a unitary mixing matrix $U_L$: 
\be \label{eq:uldef}
U_L^T \, m_\nu \, U_L = {\rm diag} (m_1,m_2,m_3)~,
\ee
where $m_i$ are the light neutrino mass eigenvalues which determine
the solar and atmospheric mass squared differences, \dms and $\dma\!\!$,
respectively. 
$M_R$ can be diagonalized by a unitary mixing matrix $U_R$ leading 
to           
\be \label{urdef}
U_R^T \, M_R \, U_R = {\rm diag}(M_1, M_2, M_3)~.
\ee
The triplet induced Majorana mass matrices in (\ref{mnu})
have the same coupling matrix $f$ in the flavor basis. Therefore, we have 
a simple relation between the left-- and right--handed masses, 
\be \label{mlmr}
\bad
M_L = f \, v_L & \mbox{ and } & M_R = f \, v_R  ~.
\ea
\ee
In (\ref{mlmr}), $v_{L, R}$ are the vevs of the left-- and right--handed
Higgs triplets, whose existence ensures the left--right
symmetry. Generically, we can translate this vevs to an approximate 
equality \cite{LR}, 
\be \label{vlvr}
v_L \, v_R \simeq \gamma \, v^2 ~,
\ee
where $v \simeq 174$ GeV is the weak scale and 
the constant $\gamma$ is a model dependent parameter of
$\cal{O}$(1). Using (\ref{mlmr}) and (\ref{vlvr}) in
(\ref{mnu}), the light neutrino mass matrix can be written as 
\be \label{mnulr}
m_\nu = v_L \,
\left( f - \tilde{M}_D \,
\frac{f^{-1}}{\gamma \, v^2} \, \tilde{M}_D^T \right) ~.
\ee                                                  
An interesting property of the mixing matrices $U_L$ and 
$U_R$, which arises due to the LR symmetry, has been 
found in \cite{JPR2}. If we assume that
$\tilde{M}_D$ is not identified with the up quark mass matrix, 
then the second term in (\ref{mnulr}) 
can be neglected and $m_\nu \simeq M_L$. Under this 
circumstance, we have $U_R \simeq U_L$, where the approximation is
true up to ${\cal{O}}(M_D^2/v^2)$ in $m_\nu$.\\

Furthermore, the approximate equality of $U_L$ and $U_R$ leads to an 
interesting and simple connection between the light and heavy 
mass eigenvalues. From $m_\nu \simeq M_L$, it follows due to (\ref{mlmr}) that 
$m_\nu \simeq \frac{v_L}{v_R} M_R$. Therefore,  
one arrives at a very simple connection between the left-- and right--handed 
sectors:
\be \label{cons}
m_i = M_i \, \frac{v_L}{v_R} ~. 
\ee
Note that in (\ref{cons}), the light neutrino masses are
proportional to the heavy right--handed masses. In other words, the low energy
spectrum is directly correlated to the spectrum at the see--saw scale. 
As we shall see 
in the next section, due to (\ref{cons}), the baryon asymmetry turns out to be 
proportional to the lightest mass eigenvalue, $m_1$.\\

In the following, we specify 
the strengths of $v_{L,R}$ which determine the corresponding size of
the light neutrino masses. From terrestrial neutrino experiments, 
the scale of the mass matrix is 
$m_\nu = v_L \, f \simeq (10^{-2} \ldots 10^{-3})$ eV\@, 
which for not 
too small $f$ is only compatible with  $v_L\, v_R \simeq \gamma \, v^2$ 
for $v_R \simeq (10^{14} \ldots 10^{15})$ GeV.
This implies that $v_R$ is probably close to the 
grand unification scale and $v_L$ is of the order of the neutrino masses.
This situation is expected since under our assumption for the Dirac mass,
$M_L$ is the dominating contribution to $m_\nu$. We shall work with 
(\ref{cons}) and explore its consequences on leptogenesis and 
low energy observables. 

\section{Leptogenesis in left--right symmetric models}
The observed baryon asymmetry, usually given as a ratio of the baryon to 
photon number density in the universe requires physics beyond the
standard model. This asymmetry can be generated by the leptogenesis 
mechanism through the mediation of sphalerons in the intermediate 
states \cite{sphaleron}. 
Within the framework of the see--saw mechanism, the heavy right--handed 
fields can produce a lepton asymmetry in an out--of--equilibrium decay.
A lepton asymmetry is caused by the interference of tree level with 
one--loop corrections to the decays of the lightest Majorana states, 
$N_1 \ra \Phi \, l^c$ and $N_1 \ra \Phi^\dagger \, l$. The resulting decay 
asymmetry reads 
\bea \label{eq:eps}  
\varepsilon = \frac{\D \Gamma (N_1 \ra \Phi \, l^c) -
\Gamma (N_1 \ra \Phi^\dagger \, l)}{\D \Gamma (N_1 \ra \Phi \, l^c) +
\Gamma (N_1 \ra \Phi^\dagger \, l)} \\[0.3cm]
= \frac{\D 1}{\D 8 \, \pi \, v^2} \frac{\D 1}{\D (M_D^\dagger M_D)_{11}}
\D\sum\limits_{j=2,3} {\rm Im} (M_D^\dagger M_D)^2_{1j} \, f(M_j^2/M_1^2)~. 
\eea
Here, $\varepsilon$ is now a function of $M_D = \tilde{M}_D U_R$ 
and the function $f$ represents the terms arising from vertex and 
self--energy 
contributions and is given to be
\be \label{fapprox}
f(x) = \sqrt{x} \left(1 + \frac{1}{1 - x} -
(1 + x) \, \ln \left(\frac{1 + x}{x}\right) \right)
\simeq - \frac{3}{2 \, \sqrt{x}}  ~.
\ee
The approximation in (\ref{fapprox}) holds for $x \gg 1$ with 
$x\equiv M_j^2/M_1^2$. It is worth
mentioning that the hierarchical assumption $(x\gg 1)$ is also favored 
in order to produce large lepton mixings within the 
see--saw mechanism \cite{smirnov}. As a 
result of (\ref{cons}), when the see--saw spectrum is hierarchical, so is 
the low energy spectrum. 
The decay asymmetry $\varepsilon$ is related to the baryon asymmetry $Y_B$ 
through the relation 
\be \label{YBth}
Y_B = c\, \kappa \frac{\varepsilon}{g^\ast}~.
\ee
In (\ref{YBth}), $c \simeq -0.55$ is the fraction of the lepton asymmetry 
converted into a baryon asymmetry via sphaleron 
processes \cite{sphaleron}, 
$\kappa$ is a suppression factor due to lepton number 
violating wash--out processes 
and $g^\ast \simeq 110$ is the number of massless degrees of freedom at 
the time of the decay. In supersymmetric models, 
$g^\ast $ and $\varepsilon$ are roughly twice as 
large, therefore the results are rather unaffected by the presence of 
supersymmetry. We shall work with the non--supersymmetric version of the 
theory. Phenomenologically, the preferred range for the baryon 
asymmetry is $ Y_B \simeq (0.6 \ldots 1) \cdot 10^{-10} $
\cite{citeYB}.\\ 

In order to estimate the baryon asymmetry, we can insert in the neutrino 
mass matrix (\ref{cons}) any of the solar solutions, i.e.\ the small 
angle (SMA), large angle (LMA) or quasi--vacuum (QVO) solution. 
Following this, the baryon asymmetry is obtained using (\ref{eq:eps}) and 
(\ref{YBth}). As a passing remark, we wish to mention 
that within the context of the left--right models, this procedure can also be
useful to extract the possible structure of the high scale theory based on the 
available phenomenological information at the low scale. This also 
relaxes the need to make, sometimes unavoidable, 
assumptions on the various neutrino parameters in order to satisfy the 
observed baryon asymmetry \cite{others1}. In addition, as we shall see, 
the contributions due to the solar and atmospheric sectors to the asymmetry
can be analyzed individually.\\

By performing a numerical analysis of the allowed oscillation 
parameters \cite{carlos} and the three unknown phases in $U_L$, 
it is found that if $\tilde{M}_D$ is a down quark or 
lepton mass matrix, $m_1$ should not be too small \cite{JPR2}. Furthermore,
the LMA solution gives a better fit to the baryon asymmetry and is thus  
slightly favored over SMA and QVO. It is interesting to note that 
current neutrino data also prefers the LMA solution \cite{krastev} over the 
other possible solutions. If $\tilde{M}_D$ is an up quark mass matrix, some 
fine tuning of the parameters is required \cite{ichAPP}.\\ 

Let us parameterize the mixing matrix $U_L$ to be of the form
\bea \label{eq:Upara}
U_L = U_{\rm CKM} \cdot P = U_{\rm CKM} \;
\cdot {\rm diag}(1, e^{i \alpha}, e^{i (\beta + \delta)}) \\[0.3cm]
= \left( \bad
c_1 c_3 & s_1 c_3 & s_3 e^{-i \delta} \\[0.2cm]
-s_1 c_2 - c_1 s_2 s_3 e^{i \delta}
& c_1 c_2 - s_1 s_2 s_3 e^{i \delta}
& s_2 c_3 \\[0.2cm]
s_1 s_2 - c_1 c_2 s_3 e^{i \delta} &
- c_1 s_2 - s_1 c_2 s_3 e^{i \delta}
& c_2 c_3\\
               \ea   \right)
 \cdot {\rm diag}(1, e^{i \alpha}, e^{i (\beta + \delta)})~,
\eea
where $c_i = \cos\theta_i$, $s_i = \sin\theta_i$ and the 
two Majorana phases are factored 
out in a matrix $P$. 
Within this parameterization, $CP$ violation in neutrino oscillations 
is governed by the Dirac phase $\delta$ and 
the $ee$ element of $m_\nu$ ($\langle m \rangle$), 
as measurable in \onbb, depends on the Majorana phases 
$\alpha$ and $\beta$. The presence and relevance of these 
additional phases was introduced in \cite{scheval1}.
If $s_3 = 0$, then the there is no $CP$ violation 
in oscillations and \meff is only a function of $\alpha$. 
In oscillation experiments, any $CP$ violation will depend on the Jarlskog
invariant \cite{jarlskog}
\be \label{JCP}
J_{CP} = \frac{1}{8} 
\sin 2 \theta_1 \sin 2 \theta_2 \sin 2 \theta_3 \cos \theta_3 
\sin \delta \lsim \frac{1}{4} s_3 c_3^2 \sin \delta ~. 
\ee 
For approximately bimaximal mixing, with 
$c_1^2 = c_2^2 = 1/2$ and keeping the leading order in $s_3$, the 
baryon asymmetry is given to be \cite{JPR2} 
\be \label{estYBLM}
Y_B \cdot 10^{10} \simeq 
\frac{4 m_1}{1 - 2 s_3 c_\delta}\left( \frac{\D m}{\D \rm GeV}\right )^2 
\left\{ \frac{s_{2\alpha} + 4 \,s_3 \, s_{\delta} \, c_{2\alpha}}
{\D \sqrt{\dms}}
+ \frac{2 s_{2(\beta + \delta) }- 4 \, s_3\, s_{2\beta+\delta}}
{\sqrt{\dma}}\right\}~. 
\ee
Here $c_\delta = \cos \delta$, $s_{2 \alpha} = \sin 2 \alpha$ and so on. 
The largest entry in $\tilde{M}_D$ is denoted by $m$.
A few remarks are in order from (\ref{estYBLM}). 
This form holds for both, the LMA and the QVO solution and clearly separates
out the contributions due to the solar and atmospheric sectors. Also, the 
Majorana phases $\alpha$ and $\beta$ do not mix and are related to the 
solar and
atmospheric sector, respectively. This feature will help us to individually
analyze the phases depending on the scales, \dms and $\dma\!\!$. It is 
explicitly seen that the baryon asymmetry 
vanishes if $CP$ conservation holds, which is the case when all 
the phases are zero or $\pi$. The asymmetry is proportional 
to the square of the heaviest entry in $\tilde{M}_D$, which in our case is
either the the tau or bottom quark mass. Due to the mass relation in
(\ref{cons}), $Y_B$ is proportional to the lightest neutrino mass 
eigenstate $m_1$ and has a lower limit of 
${\cal O}(10^{-7} \ldots 10^{-8})$ eV \cite{JPR2}. Choosing the 
Dirac mass matrix to be the up quark mass matrix will erase this 
simple proportionality of the baryon asymmetry. 
In the
following section, we use this proportionality, $Y_B \propto m_1$ to impose
a restriction following from unitarity. 
We then discuss the implications 
of this for leptonic $CP$ violation and $0\nu\beta\beta$.
To derive (\ref{estYBLM}), we assumed that the wash--out factor 
$\kappa$ is approximately 0.1.\\ 

Note that in (\ref{estYBLM}), the baryon asymmetry is predominantly
governed by the solar scale. 
However, if there are any accidental cancellations in 
the first term in (\ref{estYBLM}), then $Y_B$ will depend on the 
atmospheric scale. Alternatively, both the solar and atmospheric sectors
can contribute to the asymmetry when the two terms in (\ref{estYBLM}) are of 
the same order, which requires 
\be
\frac{\sqrt{\dma}}{\sqrt{\dms}}\simeq
\frac{2 s_{2(\beta + \delta) }- 4 \, s_3\, s_{2\beta+\delta}}
{s_{2\alpha} + 4s_3s_\delta c_{2\alpha}}\gg 1~.
\label{equa1}
\ee
This is possible when 
$s_{2\alpha} + 4 \,s_3 \, s_{\delta} \, c_{2\alpha}
\simeq 0$ or equivalently, $t_{2\alpha} \simeq -4s_3s_\delta$, where 
$t_{2\alpha} = \tan 2 \alpha$.
Given the strong 
constraints from reactor based experiments like
CHOOZ \cite{chooz} and Palo Verde \cite{paloverde} we have $0\leq 
s_3 \lsim 0.28$. Hence the atmospheric scale can contribute only if $\alpha$
is in the range such that, $-1.12 s_\delta \leq t_{2\alpha}\leq 0$. 
Clearly, this relation is not valid
for values of $\alpha \simeq (2n+1)\pi/4$.  
For example, with $\alpha \simeq n\pi/2$, we require $\delta \simeq n\pi$ in 
order to let the second term in (\ref{estYBLM}) dominate, which implies
$J_{CP}\simeq 0$. This would be identical to a two flavor scenario where we
can set $\delta =0$. An interesting outcome is that the value of $\alpha$ is
determined by the amount of $CP$ violation in oscillation experiments,
$J_{CP}$. 
Therefore, in terms of $J_{CP}$ from (\ref{JCP}), 
the atmospheric scale contributes to 
(\ref{estYBLM}) only if $\alpha$ satisfies the relation 
\be
t_{2\alpha} \simeq \frac{16J_{CP}}{(s_3^2-1)}\simeq -16J_{CP}~.
\label{atmr}
\ee
The approximation in (\ref{atmr}) is assuming that $s_3^2 \ll 1$ and this 
allows us to directly probe $\alpha$, up to ${\cal{O}}(s_3^2)$, 
by measuring the 
amount of $CP$ violation in oscillation experiments. 
In the next section, we independently analyze the 
contributions of both the solar and atmospheric sector for (\ref{estYBLM}). 
We find that given the large mass scale involved for the atmospheric sector,
it is hard to make a direct estimate of its contribution to the baryon
asymmetry.

\section{\label{sec:unit}Unitarity bound, 
neutrinoless double beta decay and $CP$ violation}
In the presence of any new physics originating at some scale $M_X $ 
above the electroweak scale, one  
should consider the standard model as an effective theory. An upper limit 
for $M_X$ can be determined by examining the high energy behavior of the 
lepton number violating reactions like $\nu\nu\to WW$ or $ZZ$, 
which can occur 
because of a Majorana mass term. It was noted that a 
stringent bound for $M_X$ is obtained by considering the following 
linear combination of the zeroth partial wave amplitudes \cite{maltoni}
\[ 
a_0\left(\frac{1}{2}(\nu_{+}\nu_{+}-\nu_{-}\nu_{-})\to\frac{1}{\sqrt{3}}
(W^+ W^- +Z^0 Z^0 ) \right) \, , 
\]
where $\nu_{\pm}$ are helicity components 
of the neutrino mass eigenstate and the final state bosons are longitudinally 
polarized. This amplitude to obey unitarity requires 
$\vert a_0\vert\leq 1/2$. In terms of the mass eigenvalue $m_1$, this 
can be translated to \cite{maltoni}
\be
m_1 = \frac{4\pi v^2}{\sqrt{3} M_X}~.
\label{bound1}
\ee
In the left--right symmetric model, the Higgs triplet  
$\Delta_L = (\Delta^0, \Delta^+, \Delta^{++})$, which generates the light 
left--handed neutrino masses, also induces lepton number violating processes 
like $e^+ e^+ \ra W^+ W^+$. The $t$ and the $u$ channels of this process 
exhibit a unitarity violating high energy behavior, since the amplitude 
grows with energy. It has been explicitly shown that the presence of the 
doubly charged Higgs boson $(\Delta^{++})$, which mediates the same 
process via exchange in the $s$ channel, restores unitarity in the
high energy limit \cite{eeWW}. Furthermore, it follows from (\ref{vlvr}), 
that for a light $m_1$, the scale $M_X$ is dependent on $\gamma$ and for 
$\gamma \simeq 1$ we require $M_X \sim v_R$. However, in the following, 
for our analysis, it is sufficient if we maintain the requirement that 
$M_X$ be lower than the Planck scale.\\

Using (\ref{estYBLM}), we have an equality relating the scale $M_X$ 
to  $Y_B$: 
\be \label{eqn1}
M_X \simeq \frac{\D 16\pi v^2}{\sqrt{3}
\D Y_B \cdot 10^{10}(1-2s_3c_\delta)} \left( \frac{\D m}{\D \rm GeV}\right)^2 
\left\{\frac{s_{2\alpha} + 4 \,s_3 \, s_{\delta} \, c_{2\alpha}}{\sqrt{\dms}}
+ \frac{2 s_{2(\beta + \delta) }- 4 \, s_3\, s_{2\beta+\delta}}
{\sqrt{\dma}}\right\}~. 
\ee
As observed in the previous section, there are contributions due to the
solar and atmospheric sector. We analyze them separately.

\subsection{Effects due to the solar scale}
In the following, we neglect the contribution due to the atmospheric scale,
and examine (\ref{eqn1}) for the LMA and QVO 
solar solutions. In this case, the Majorana phase $\beta$ is a free 
parameter in the theory. Depending on the solar solution, constraints on 
$\alpha$ and $\delta$ are obtained, which could reflect in low energy 
observables such as \meff and in the $CP$ violating parameter, $J_{CP}$. 
In order to satisfy the baryogenesis requirement, we set 
$Y_B = 10^{-10}$ and (\ref{eqn1}) can be rewritten as  
\be
M_X \simeq  \left( \frac{\D m}{\D \rm GeV}\right)^2 
\frac {16 \pi v^2  }{\D \sqrt{3\dms}} A  \; \mbox{, where } \; A =
\frac{s_{2\alpha} + 4 \,s_3 \, s_{\delta} \, c_{2\alpha}}{1-2s_3c_\delta}
~.
\label{eqn2}
\ee

The lower limit of $s_3 = 0$ is identical to a two flavor system, where one
can set the Dirac $CP$ phase to zero. 
The fact that in a two flavor limit there is still a $CP$ violating phase 
($\alpha$) reflects the Majorana nature of the neutrinos involved.
Choosing $\tilde{M}_D$ to be the charged lepton mass matrix, therefore  
$m = m_\tau = 1.77$ GeV\@, we have
\be 
M_X \simeq 2.8 \cdot 10^{15}\frac {A}{\D \sqrt{\dms}}\mbox{ GeV} 
\leq M_{\rm Pl}
= 1.2 \cdot 10^{19} \mbox{ GeV}~,
\label{eqn3}
\ee
where \dms is given in eV$^2$ and 
$M_{\rm Pl}$ denotes the Planck scale. This sets an upper bound on 
$A$ for a given $\dms\!\!$. Furthermore, it could restrict the values 
in \meff and $J_{CP}$. This approach can be useful to probe the possible 
value for $M_X$ based on our chosen low energy observables.
Future terrestrial solar experiments like BOREXINO \cite{borexino}, which 
will identify the preferred \dms$\!\!$, can also correlate to the scale in our
scheme. However, we can still make an estimate of the size of $\dms\!\!$, 
depending on the parameters. To see this, we rewrite the result 
in $(\ref{eqn3})$ as
\be
A \leq 4.3\cdot 10^3\sqrt{\dms}/\mbox{eV}~.
\label{eqn4}
\ee
Thus, depending on the values for $\alpha$, $\delta$ and for 
a given $s_3$, a lower bound on $\sqrt{\dms}$ is possible. For example, the 
maximum value that $A$ can take is for the case when $\delta=0$ and  
$s_3=0.28$, which is its maximally allowed value. For this choice, we have 
$max(A) \simeq 2.3$. Correspondingly, this sets a lower bound of  
\be
\dms \!\!\geq 2.7\cdot 10^{-7}~\mbox{eV}^2~\mbox{for}~A=2.3~.
\label{eqn5}
\ee 
Note that from $(\ref{eqn5})$, in order to incorporate the QVO solar
solution, we need to restrict the value of $A$ much below its upper limit. 
As we shall see in the following, the allowed region of parameter space
for the LMA solution clearly covers $A\geq 2.3$ and for the QVO solution 
the parameter space is restricted with $A\ll 2.3$. 
 
\subsubsection{LMA solution}
In this case, we choose $\dms\!\! \simeq 5\cdot 10^{-5}$ eV$^2$ 
and for this value we have $M_X \simeq 3.9 \cdot 10^{17} A$ GeV\@. 
Following (\ref{eqn3}) we can have a closed bound 
\be
0 \leq \frac{s_{2\alpha} + 4 s_3 s_\delta c_{2\alpha}}{1 - 2 s_3 c_{\delta}}
\lsim 30.6~.
\label{lmabound1}
\ee
Clearly, there are no restrictions in the various angles, $\alpha$ and 
$\delta$ in order to satisfy the bound in (\ref{lmabound1}). 
Note that for the two flavor limit, the bound in (\ref{eqn4}) gives a 
consistent upper bound for one of the Majorana phases, 
\be
s_{2\alpha}\leq 4.3\cdot 10^3\cdot \sqrt{\dms}/\mbox{eV}~\mbox{or}~
\dms \! \geq 5.4\cdot 10^{-8}\mbox{eV}^2~\mbox{for}~\alpha=\pi/4~.
\label{eqn6}
\ee
This bound can however be revised 
if we lower the scale where unitarity may break down. For LMA, one 
requires $m_1 \lsim 10^{-3}$ eV in order to have a hierarchical scheme, 
which corresponds to $M_X \gsim 2.2 \cdot 10^{17}$ GeV\@. 
Then, (\ref{lmabound1}) is modified to the range 
\be
0.6\leq \frac{s_{2\alpha} + 4 s_3 s_\delta c_{2\alpha}}{1 - 2 s_3 c_{\delta}}
\lsim 30.6~,
\label{lmabound2}
\ee
which can be used to set bounds on the phases.
On the other hand, for the hierarchical scheme to hold in the QVO solution, 
one needs values of $M_X$ close to the Planck scale. 
In Fig.\ \ref{aldel5}, 
we show the area in $\alpha$--$\delta$ space which is 
allowed for $\dms \!\! \simeq 5 \cdot 10^{-5}$ eV$^2$ and $s_3^2 = 0.08$ and 
0.001, respectively. 
As expected, the allowed
region for the Majorana phase $\alpha$ is strongly constrained while the
$CP$ violating phase $\delta$ remains unbounded. This is an indication of
possible effects in \meff while $J_{CP}$ could still not be sensitive to
our constraints.\\

From Fig.\ \ref{aldel5} one observes that the phase $\alpha$ is basically 
around $\pi/4$ or $5\pi/4$, which incidentally, from (\ref{atmr}), are the 
values disallowed when the atmospheric scale contributes to the asymmetry. 
If $\alpha$ is fixed, then \meff is a function of the second phase $\beta$. 
As known, the LMA solution provides the highest value for \meff in 
the hierarchical scheme. 
For $\alpha = \pi/4$ we show in Fig.\ 
\ref{meeLMA1} the expected \meff for different \dms and $s_3^2$. 
For $\alpha = 5\pi/4$ the situation is basically the same. 
A measurement of \meff could probe the second phase 
$\beta$, which drops out of the baryon asymmetry in our scenario. 
The limiting values are 
$0.001 \mbox{ eV} \lsim \meff \!\!  \lsim 0.01\mbox{ eV}$,  
depending on the values for \dms and $s_3$. Large part of this 
range is well within the sensitivity of the GENIUS experiment \cite{GENIUS}. 
If $s_3^2$ is too small, the dependence on $\beta$ vanishes, as does 
the presence of $CP$ violation in oscillation experiments. 
We remark that one could in principle obtain all phases by measuring 
the other entries of the light neutrino mass matrix, 
e.g.\ the element $m_{\mu\mu}$, 
which triggers the decay $K^+ \ra \pi^- \mu^+ \mu^+$. However, this 
and other analogue processes have far too low branching ratios to be 
observed \cite{ichPRD}. 

\subsubsection{QVO solution}
In this case, requiring $M_X \le M_{\rm Pl}$ and for  
$\dms \!\! \simeq 5 \cdot 10^{-10}$ eV$^2$, this corresponds to
 $M_X \simeq 1.2 \cdot 10^{20} A $ GeV. Similar to the bound in 
(\ref{lmabound1}), we now have 
\be
0 \leq \frac{s_{2\alpha} + 4 s_3 s_\delta c_{2\alpha}}{1- 2 s_3 c_{\delta}}
\lsim 0.1~.
\label{qvobound}
\ee
This is a stronger limit than the one obtained for the LMA solution. It 
also imposes restrictions on the values for $\alpha$, 
$\delta$ and $s_3$, and requires the value of 
$A$ to be lower than the upper limit 
suggested in (\ref{eqn5}). In Figs.\ \ref{aldel8} and \ref{aldel10}, 
we plot the allowed areas in $\alpha$--$\delta$ space 
for different values for $s_3$ and \dms$\!\!$.
As can be seen from these figures, in contrast to the LMA solution,
the phase $\alpha$ is basically around $\pi/2$ or $\pi$. 
For these two choices of $\alpha$ we show in Figs.\ 
\ref{meeQVO1} and \ref{meeQVO2} the value of \meff for different \dms 
and $s_3^2$. The limiting values are now 
$10^{-5}\mbox{ eV} \leq \meff \!\! \leq 0.01\mbox{ eV}$,  
depending on the values for \dms and $s_3$. Some part of this 
range is well within the sensitivity of the GENIUS experiment.  
As known, for the QVO solution large $s_3^2$ is required in order 
to give accessible \meff$\!\!$. We note that the cases 
$\alpha = \pi$ or $\pi/2$ together with $\beta = \pi$ or $\pi/2$ are 
situations in which one can not distinguish $CP$ violation from $CP$ 
conservation in \obb \cite{ichNPB}.\\

A common feature for both solar solutions is that the case $\delta = 0$ 
is allowed, which can be seen from 
Figs.\ \ref{aldel5}, \ref{aldel8} and \ref{aldel10}. 
This means that vanishing $CP$ violation in oscillation 
experiments does not mean that leptogenesis is disfavored. Also, since 
the atmospheric scale decouples in our framework, the presence of only 
one non--vanishing Majorana phase is sufficient to generate the observed 
baryon asymmetry. These features have also been observed in the model 
presented in \cite{port}.

\subsection{Effects due to the atmospheric scale}
As mentioned earlier, for the contributions from the atmospheric scale to 
be significant, we require to satisfy (\ref{atmr}). This leads to 
two possibilities: case (i) with $CP$ violation or
Im$(U_{e3})\neq 0$ and case (ii) with no $CP$ violation or Im$(U_{e3})=0$.
Note that case (ii) is also identical to the two flavor scenario where we
can set $\delta =0$. However, as already hinted, due to the largeness of the 
atmospheric scale, regardless of the unitarity bound, we do not expect to 
have strong constraints on the Majorana phases, unlike the situation for 
the solar sector.\\
 
We first analyze the possibility where Im$(U_{e3})\neq 0$. 
For this analysis, we take the largest value of $s_3=0.28$, for which
case we can express the scale $M_X$ as 
\be
M_X \simeq \frac{\D 16\pi v^2}{\sqrt{3\dma}
\D Y_B \cdot 10^{10}}\left( \frac{\D m}{\D \rm GeV}\right)^2 
\tilde A \leq M_{\rm Pl}~\mbox{, where}~
\tilde A = \frac{2 s_{2(\beta + \delta) }- 1.12\, s_{2\beta+\delta}}
{1-0.56c_\delta}~. 
\label{matm}
\ee
As in the previous cases, setting $Y_B = 10^{-10}$ and for $m =m_\tau =
1.77$ GeV, we have the lower bound
\be
\sqrt{\dma}\geq 2.29\cdot 10^{-4}\tilde A ~\mbox{eV}~,
\label{atb1}
\ee
which is easily satisfied for any value of the angles, 
$\beta$ and $\delta$. Furthermore, setting $\dma \simeq 
3\cdot 10^{-3}$eV$^2$, we have the closed bound
\be
0\leq \frac{2 s_{2(\beta + \delta) }- 1.12\, s_{2\beta+\delta}}
{1-0.56c_\delta}\leq 187.6 
\label{atb2}
\ee
and there are no restrictions on the angles from (\ref{atb2}). The reason
for this uninteresting situation is that the atmospheric scale is too large 
to set any useful limits on $\tilde A$. As a result, the bounds have no impact
on either \meff or $J_{CP}$. In case (ii), where we have no $CP$ violation, 
we set $\delta =0$ in all the results obtained above for the $CP$ violating 
scenario. The only Majorana parameter $\beta$, as expected, satisfies all 
the bounds derived above and remains unconstrained. As a further check,
we briefly address the question: could there be any restrictions on $\beta$
if we relate the estimates from \obb and $J_{CP}$? To see this, 
in the hierarchical scheme of neutrino masses, we have 
\be
\meff \!\! \simeq 
\left[\dms s_1^4 + \dma s_3^4 +2 s_1^2 s_3^2 \sqrt{\dms \dma} c_{\phi}
\right]^{1/2}~, 
\label{meffatm}
\ee
where $\phi = 2(\alpha - \beta)$. 
As a numerical illustration, we choose
$\dms \!\! = 5\cdot 10^{-5}~\mbox{eV}^2,~\dma \!\! 
= 5\cdot 10^{-3}~\mbox{eV}^2,~
s_3 = 0.28~\mbox{and}~s_1 = \pi/4$, for which case
\be
\meff \!\! \simeq 6.3 \cdot \sqrt{1 + c_{\phi}}\cdot 10^{-3}~\mbox{eV}~.
\label{numeffatm}
\ee
Using (\ref{atmr}), we can rewrite (\ref{numeffatm}) as
\be
\meff \!\! \simeq 6.3 \left(1+\cos \left[\tan^{-1} 16J_{CP} + 2 
\beta \right] \right)^{1/2}\cdot 10^{-3}~\mbox{eV}.
\label{cpatm}
\ee
Taking the current allowed range, $0\leq \meff \!\! \leq 0.35$ eV, this 
translates to a closed bound 
\be
0\lsim \cos\left[\frac{1}{2}\tan^{-1} 16J_{CP} + \beta\right] \lsim 39.2~, 
\label{cpr}
\ee
which, as expected, 
is easily satisfied for all $\beta$ and $J_{CP}$. We see 
from (\ref{numeffatm}) that almost all of the allowed regions are well within
the reach of GENIUS except for $\phi = \pi$ for which case we arrive at 
$\tan2\beta \simeq -16J_{CP}$. Therefore, for a null 
\obb result, the Majorana phase $\beta$ is determined by the amount of
$CP$ violation in oscillation experiments. Again, this conclusion requires 
some finetuning, now in \meff$\!\!$, and is equivalent to the requirement in 
(\ref{atmr}).\\

\section{\label{sec:sum}Summary}
In left--right symmetric theories one can find a simple 
formula for the baryon asymmetry, expressing it in terms of the 
low energy neutrino parameters. In our analysis, we have made a specific 
choice for the Dirac mass matrix to be the charged lepton mass matrix. This
choice results in only the triplet term contributing to the neutrino mass, 
while, for all practical purpose, 
the conventional see--saw term gives a negligible 
contribution. This results in a simple expression for $Y_B$ which is 
proportional to the lightest Majorana mass $m_1$. On the other hand, if we
choose $\tilde{M}_D$ to be the up--quark mass matrix, 
such a proportionality is not 
possible. We find that within this model the imposition of an additional 
constraint on $m_1$ coming from unitarity restricts the allowed parameter 
space for the $CP$ violating phases. This could be a distinguishing feature 
of the choice of the Dirac mass matrix with observable low energy 
consequences. The ensuing bound helps in narrowing 
down one of the Majorana phases, thereby reducing the theoretical 
uncertainty in the prediction for \meff$\!\!$. The value of the phase 
constrained by our approach is different for the LMA and QVO solution. 
Upon measuring the effective 
mass, one could obtain the second Majorana phase. In each of the cases, the 
corresponding limit of a two flavor system is obtained by setting $s_3 =0$. 
In general, for both cases, most of the allowed parameter space  
predicts \meff in the measurable range of GENIUS with its sensitivity of 
$\meff \!\! \geq 10^{-3}$ eV. The presence of one single 
Majorana phase is sufficient to generate the correct baryon asymmetry, 
especially the case $\delta = 0$ is allowed, which corresponds to 
no $CP$ violation in oscillation experiments. 
We examined individually the
contributions to the asymmetry due to the solar and atmospheric sectors and
found that in general the solar mass scale 
dominates $Y_B$. 
Under a special situation, when 
$\meff \!\! \simeq 0$, one 
could relate the phase $\beta$ to $CP$ violation 
in oscillations experiments. This might perhaps indicate the possible
atmospheric contribution. However, if there are no positive indications 
for \meff from GENIUS, we still need to rule out the QVO solution in order to 
strengthen our claim for the atmospheric contribution. 
The reason is that the QVO 
solution predicts a small though nonzero $\meff\!\! \leq 10^{-4}$ eV
(see Figs.\ 5 and 6). 
It is in this context that future experiments like BOREXINO, which 
can pin down the correct solar solution, will help in better understanding 
the various $CP$ phases within this scenario. 
Furthermore, upon correlation with long baseline 
experiments searching for a nonzero $U_{e3}$ and $CP$ violation, 
together with 
a simultaneous measurement of \obb$\!\!$, 
it might be possible to make a reasonable 
guess on the Majorana phase contributions to the baryon asymmetry.

\begin{center} 
{\bf Acknowledgments} 
\end{center} 
This work has been supported by the
Bundesministerium f\"ur Bildung, Wissenschaft, Forschung und Technologie,
Bonn under contract no. 05HT1PEA9. W.R.\ acknowledges the 
financial support from the Graduate College ``Erzeugung und Zerf\"alle 
von Elementarteilchen'' at Dortmund University. We also thank Prof.\ R.N. 
Mohapatra and A.P. Lorenzana for useful comments and discussions.


\begin{center}

\begin{samepage}
\begin{figure}[ht]\hspace{-1.6cm}
\epsfig{file=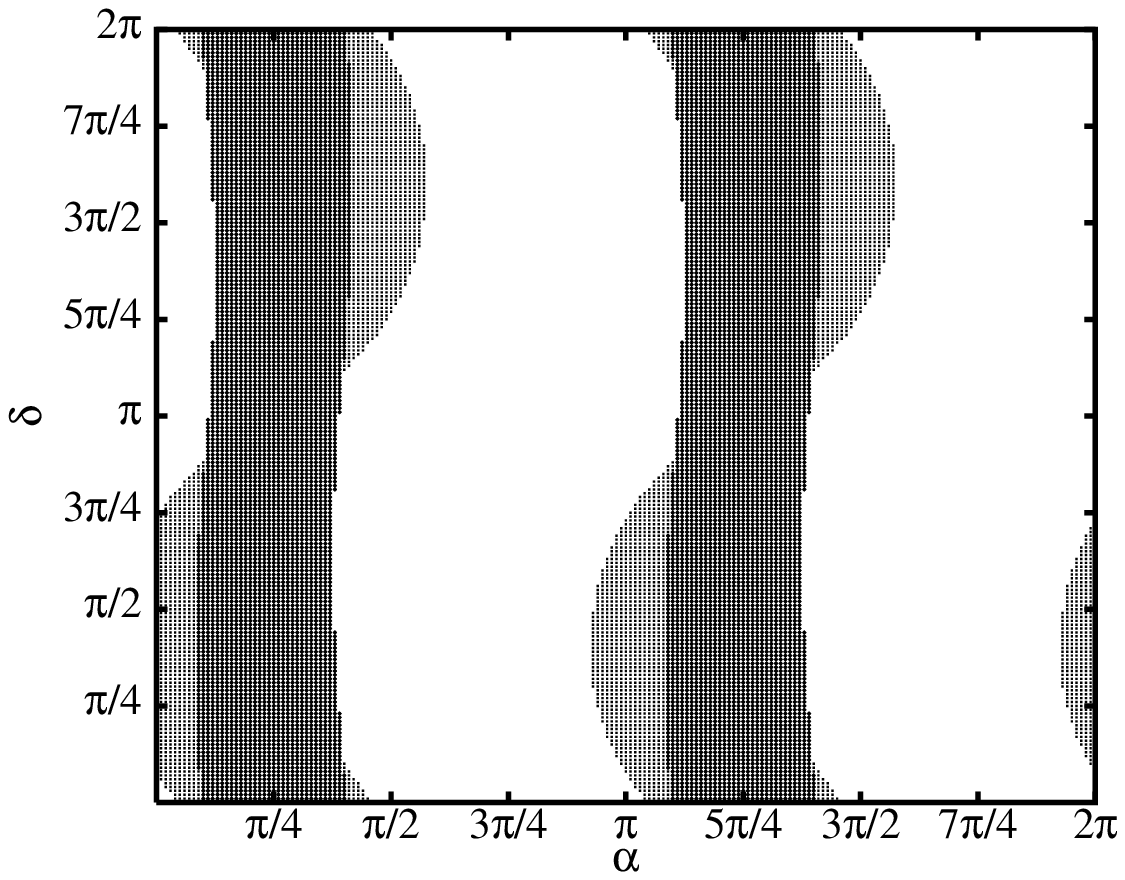,width=14cm,height=9cm}
\caption{\label{aldel5}Allowed area in $\alpha$--$\delta$ space 
in the LMA solution for $M_X = 2.2 \cdot 10^{17}$ GeV, 
$\dms \!\! = 5 \cdot 10^{-5}$ eV$^2$, $s_3^2=0.08$ (dark shaded) and 
 $s_3^2=0.001$ (light shaded).}
\epsfig{file=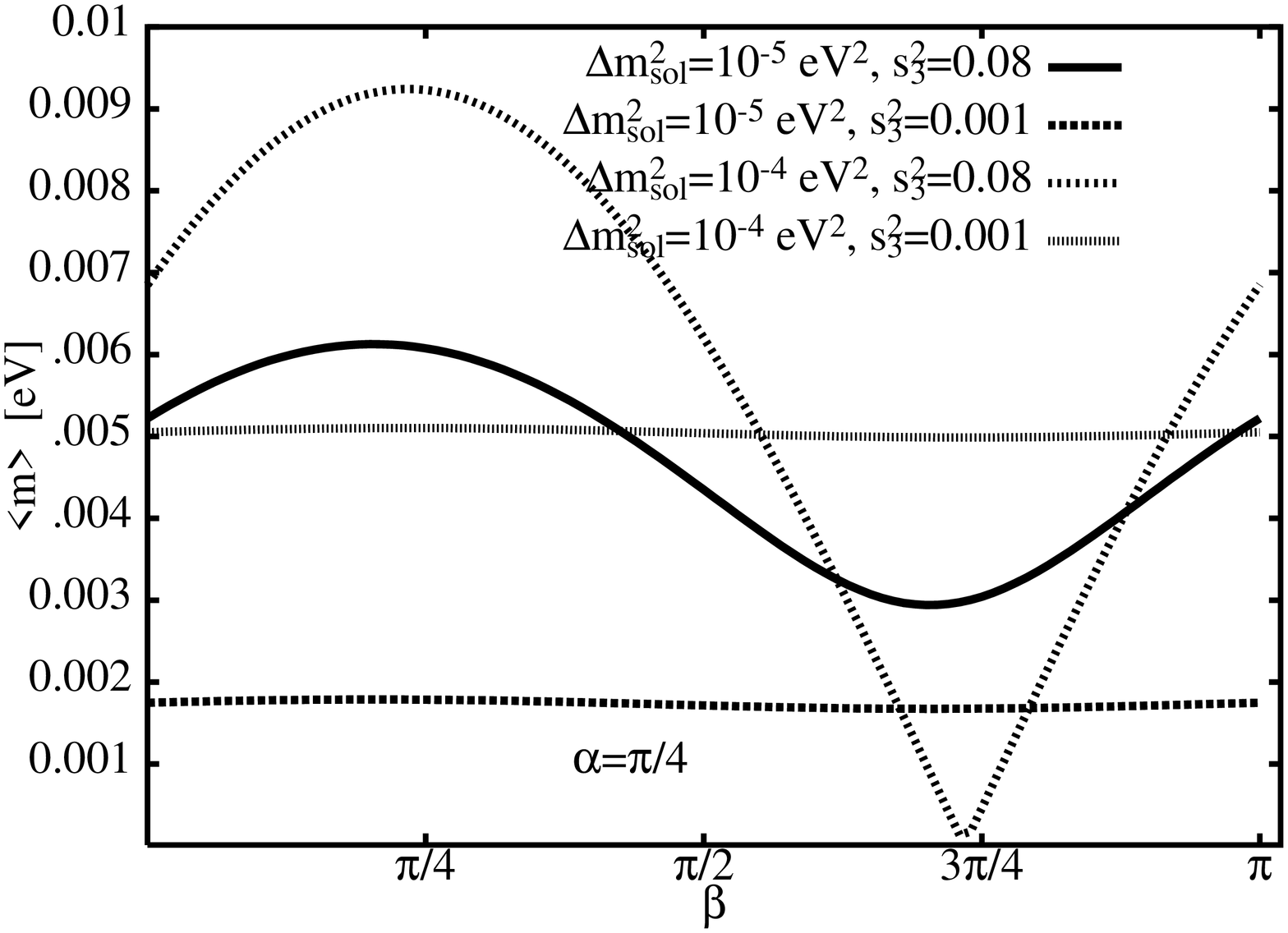,width=9cm,height=14cm,angle=270}
\caption{\label{meeLMA1}\meff 
as a function of $\beta$ for $\alpha = \pi/4$, 
different \dms and $s_3^2$.}
\end{figure} 
\end{samepage}

\begin{figure}[ht]\hspace{-1.6cm}
\epsfig{file=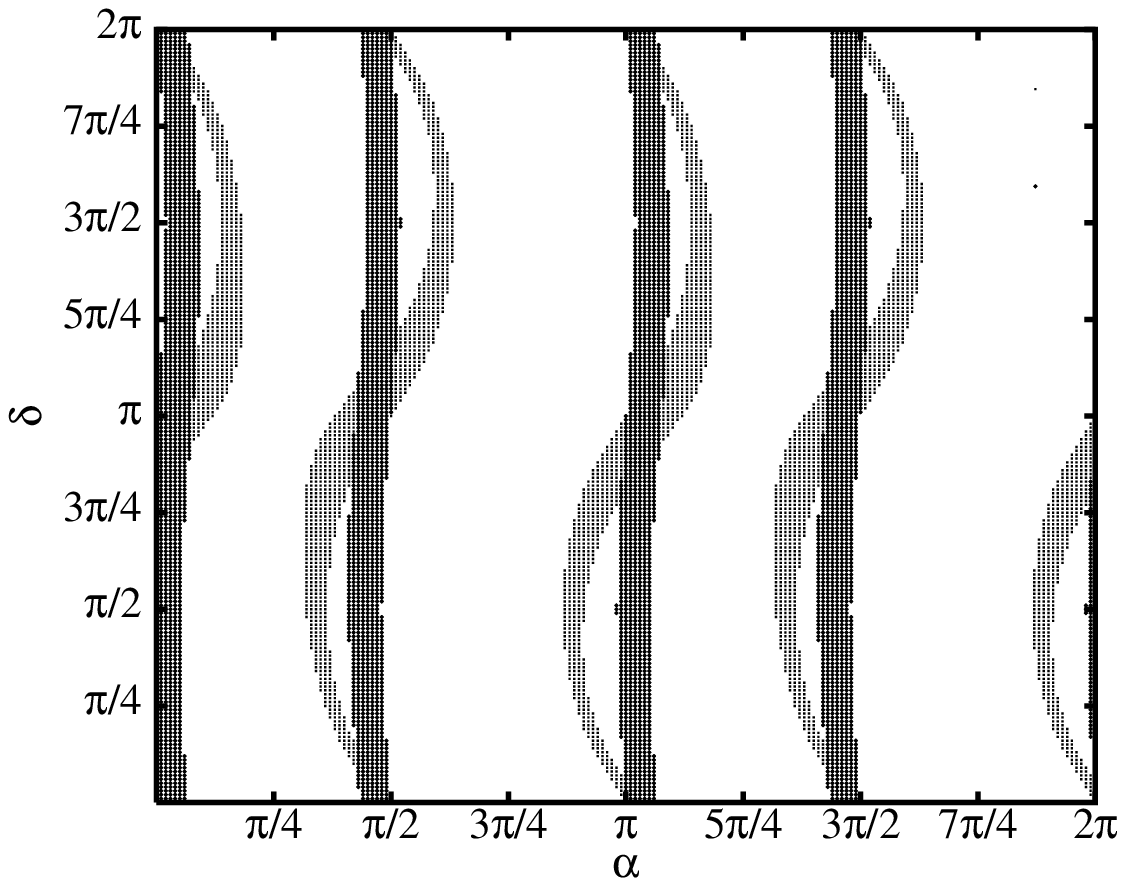,width=14cm,height=9cm}
\caption{\label{aldel8}Allowed area in $\alpha$--$\delta$ space 
in the QVO solution for $M_X = M_{\rm Pl}$, 
$\dms \!\! = 10^{-8}$ eV$^2$, $s_3^2=0.08$ (dark shaded) and 
 $s_3^2=0.001$ (light shaded).}
\end{figure}

\begin{figure}[hb]\hspace{-1.6cm}
\epsfig{file=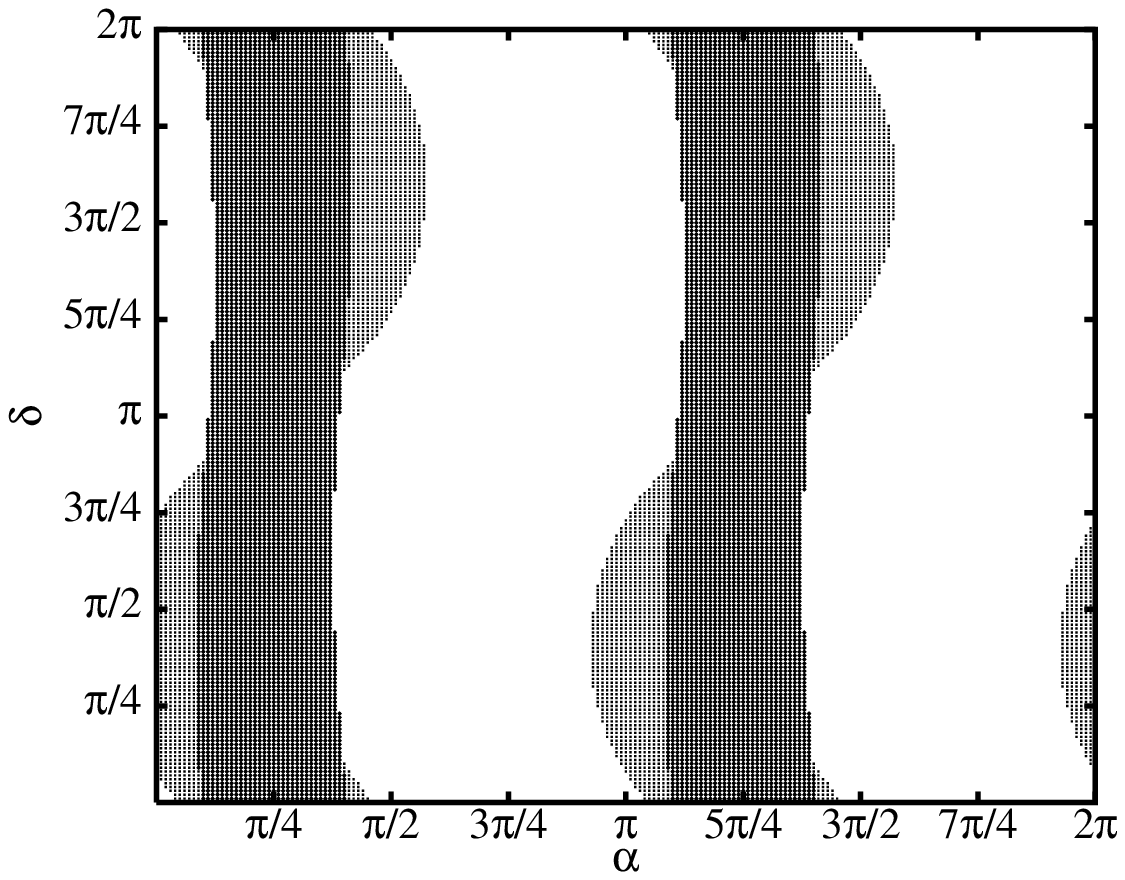,width=14cm,height=9cm}
\caption{\label{aldel10}Allowed area in $\alpha$--$\delta$ space 
in the QVO solution for 
$M_X = M_{\rm Pl}$, 
$\dms = \!\! 10^{-10}$ eV$^2$, $s_3^2=0.08$ (dark shaded) and 
 $s_3^2=0.001$ (light shaded).}
\end{figure}

\begin{figure}[ht]
\epsfig{file=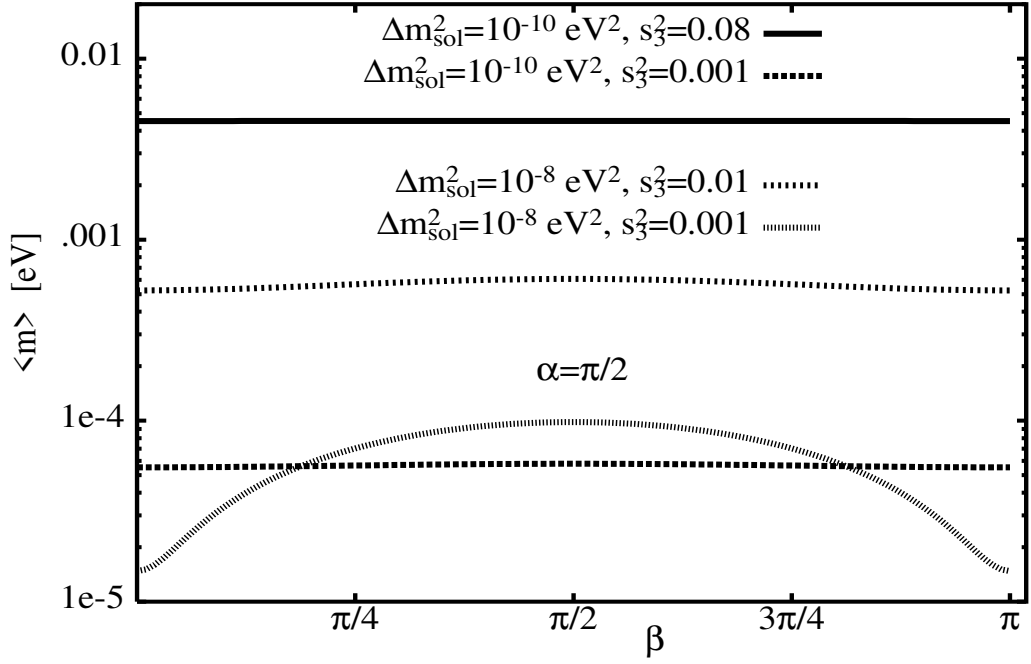,width=14cm,height=9cm}
\caption{\label{meeQVO1}\meff in the QVO solution 
as a function of $\beta$ for $\alpha = \pi/2$, 
different \dms and $s_3^2$.}
\end{figure} 

\begin{figure}[hb]
\epsfig{file=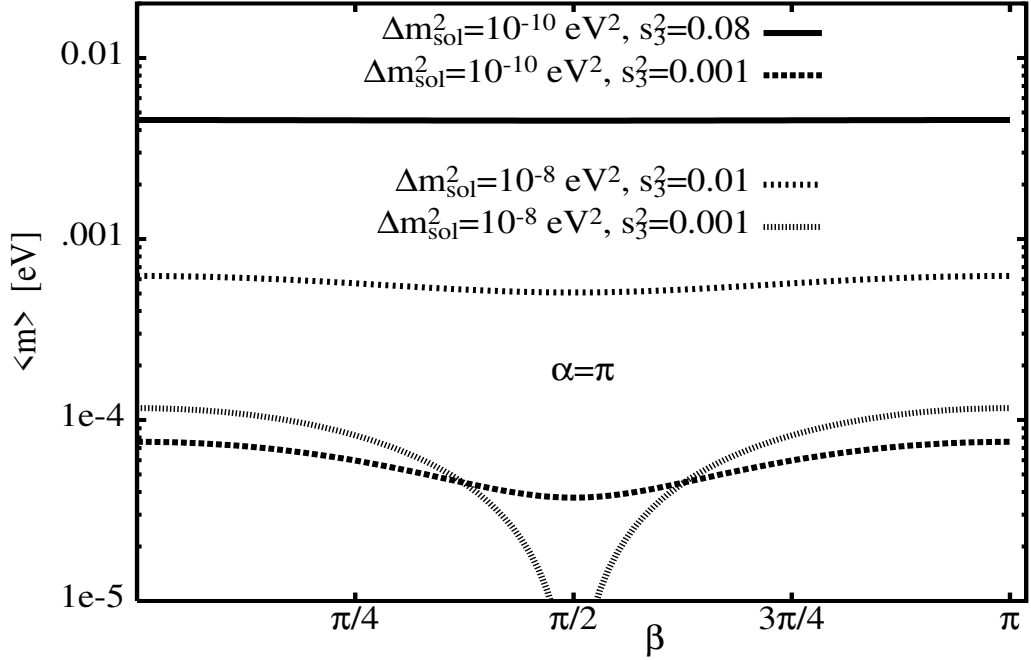,width=14cm,height=9cm}
\caption{\label{meeQVO2}\meff in the QVO solution 
as a function of $\beta$ for $\alpha = \pi$, 
different \dms and $s_3^2$.}
\end{figure} 
 
\end{center}                           

\end{document}